\documentstyle[aps,epsf]{revtex}  

\newcommand{\etal}{\textit{et~al.}}
\newcommand{\pho}{\phantom{0}}
\newcommand{\stat}{(\mathrm{stat})}
\newcommand{\syst}{(\mathrm{syst})}

\newcommand{\mrad}{\mathrm{\,mrad}}
\newcommand{\MeV}{\mathrm{\,Me\hspace{-0.1em}V}}
\newcommand{\GeV}{\mathrm{\,Ge\hspace{-0.1em}V}}
\newcommand{\pb}{\mathrm{\,pb^{-1}}}
\newcommand{\Br}{\mathrm{Br}}
\newcommand{\boldpi}{\mathbf{\pi}}

\newcommand{\Zboson}{\mathrm{Z}}

\newcommand{\Bmeson}{\mathrm{B}}
\newcommand{\boldBmeson}{\mathbf{B}}
\newcommand{\BmesonUD}{\mathrm{B}_{u,d}}
\newcommand{\Bstar}{\mathrm{B}^{*}}

\newcommand{\BorBstar}{\mathrm{B}^{(*)}}

\newcommand{\Bdstar}{\mathrm{B}^{**}}

\newcommand{\BdstarUD}{\mathrm{B}^{**}_{u,d}}
\newcommand{\BdstarS}{\mathrm{B}^{**}_{s}}
\newcommand{\Btwostar}{\mathrm{B}_2^{*}}
\newcommand{\Bone}{\mathrm{B}_1}
\newcommand{\Bonestar}{\mathrm{B}_1^{*}}
\newcommand{\Bzerostar}{\mathrm{B}_0^{*}}
\newcommand{\Dmeson}{\mathrm{D}}

\newcommand{\Dstar}{\mathrm{D}^{*}}

\newcommand{\Kmeson}{\mathrm{K}}

\begin{document}        

\baselineskip 14pt
\title{Measurement of the Spectroscopy of Orbitally Excited B~Mesons \\
  with the L3 detector}
\author{Vuko Brigljevi\'{c}\footnote{Representing the L3 collaboration.}}
\address{Institute for Particle Physics, ETH Zurich \\
CH-8093 Zurich, Switzerland}
%
\maketitle              

\begin{abstract}        
We measure the mass, decay width and production
rate of orbitally excited B mesons in $1.25$ million hadronic Z decays
registered by the L3 detector in 1994 and 1995.  B meson candidates are
inclusively reconstructed and combined with charged pions produced at the event
primary vertex.  An excess of events above the expected background is observed
in the $\Bmeson\pi$ mass spectrum near $5.7~\GeV$.  These events are interpreted
as resulting from the decay $\Bdstar \rightarrow \BorBstar\pi$, where $\Bdstar$
denotes a mixture of $L=1$ B meson spin states.  The masses and decay widths of
the $\Btwostar$ ($j_q = 3/2$) and $\Bonestar$ ($j_q = 1/2$) resonances and the
relative production rate for the combination of all spin states are extracted
from a fit to the mass spectrum. 
\end{abstract}          

\section{Introduction}               

Detailed understanding of the resonant structure of orbitally excited B 
mesons provides important information regarding the underlying theory.  A
symmetry (Heavy Quark Symmetry) arises from the fact that the mass of the $b$
quark is large relative to $\Lambda_{\mathrm{QCD}}$.  In this approximation, the
spin of the heavy quark ($\vec{s}_Q$) is conserved independently of the total
angular momentum ($\vec{j}_q = \vec{s}_q + \vec{l}$) of the light quark.
Excitation energy levels are thus degenerate doublets in total spin and can be
expressed in terms of the spin-parity of the meson $J^P$ and the total spin of
the light quark $j_q$. Corrections to this symmetry are a series expansion in 
powers of $1/m_Q$\cite{Isgur}, calculable in Heavy Quark
Effective Theory (HQET).

The $L=0$ mesons, for which $j_q = 1/2$, have two possible
spin states: a pseudo-scalar $P$ ($J^P = 0^-$)  and a vector $V$ ($J^P = 1^-$). 
If the spin of the heavy quark is conserved independently, the relative production
rate of these states is $V/(V+P) = 0.75$.\footnote{Corrections due to the decay of
higher excited states are predicted to be small.}  Recent measurements of this 
rate for the $\Bmeson$ system \cite{BstarL3,BstarDELPHI,BstarOPAL,BstarALEPH} 
agree well with this ratio.

In the case of orbitally excited $L=1$ mesons, two sets of
degenerate doublets are expected: one corresponding to $j_q = 1/2$ and the 
other to $j_q = 3/2$.  Their relative production rates follow from spin
state counting ($2J+1$ states). 
Rules for the decay of these states to the $1S$ states are determined by 
spin-parity conservation~\cite{Isgur,Rosner}.  For the dominant two-body decays, 
the $j_q = 1/2$ states can decay via an $L=0$ transition (S-wave) and their decay 
widths are expected to be broad in comparison to those of the $j_q = 3/2$ states 
which must decay via an $L=2$ transition (D-wave).  Table~\ref{tab:decays} presents
the nomenclature of the various spin states for $L=1$ $\Bmeson$ mesons containing 
either a $u$ or $d$ quark, with the predicted production rates and two-body decay 
modes.

Several models, based on HQET and on the charmed $L=1$ meson data, have made 
predictions for the masses and widths of orbitally excited $\Bmeson$ mesons.
Some of these models~\cite{Gronau,Eichten,Falk} place the average mass of the 
$j_q = 3/2$ states above that of the $j_q = 1/2$ states, while 
others~\cite{Isgur,Ebert} predict the opposite  
(``spin-orbit inversion''). 

\begin{table}[b]
  \begin{center}
    \caption{Spin states of the $L=1$ mesons with their predicted production
             rates and decay modes.}\label{tab:decays}
    \begin{tabular}{cccccc}
      Name         & $j_q$ & $J^P$ & Production &  Decay mode                       & Transition \\
      \hline
      $\Bzerostar$ & $1/2$ & $0^+$ & 1/12       & $\Bzerostar\rightarrow\Bmeson\pi$ & S-wave \\
      $\Bonestar$  & $1/2$ & $1^+$ & 3/12       & $\Bonestar\rightarrow\Bstar\pi$   & S-wave \\
      $\Bone$      & $3/2$ & $1^+$ & 3/12       & $\Bone\rightarrow\Bstar\pi$       & D-wave \\
      $\Btwostar$  & $3/2$ & $2^+$ & 5/12       & $\Btwostar\rightarrow\Bstar\pi,\Bmeson\pi$ & D-wave \\
    \end{tabular}
  \end{center}
\end{table}

Recent analyses at LEP combining a charged pion produced at the primary event
vertex with an inclusively reconstructed $\Bmeson$ 
meson~\cite{BstarALEPH,BdstarOPAL,BdstarDELPHI}
have measured an average mass of $M_{\Bdstar} = 5700-5730\MeV$, where $\Bdstar$
indicates a mixture of all $L=1$ spin states.  An analysis \cite{BdstarALEPH}
combining a primary charged pion with a fully reconstructed $\Bmeson$ meson,
measures $M_{\Btwostar} = (5739\pho^{+8}_{-11}\stat\pho^{+6}_{-4}\syst)\MeV$ by
performing a fit to the mass spectrum which fixes the mass differences, widths
and relative rates of all spin states according to the predictions of
Eichten, \etal.\cite{Eichten}.

The analysis presented here~\cite{L3Bdstar} is based on the combination of primary 
charged pions
with inclusively reconstructed $\Bmeson$ mesons.  Several new analysis techniques
make it possible to improve on the resolution of the $\Bmeson\pi$ mass spectrum
and to unfold this resolution from the signal components.  As a result,
measurements are obtained for masses and widths of D-wave $\Btwostar$ decays and
of S-wave $\Bonestar$ decays.

\section{Event Selection}

\subsection{Selection of $\Zboson \rightarrow b\bar{b}$ decays}

The analysis is performed on data collected by the L3 detector \cite{L3} in 1994 
and 1995, corresponding  to an integrated luminosity of $90\pb$ with LEP operating 
at the Z mass.  Hadronic Z decays are selected\cite{Hadrons} which have an  event 
thrust direction satisfying $|\cos\theta| < 0.74$, where $\theta$ is the polar angle. 
The events are also required to
contain an event primary vertex reconstructed in three dimensions, at least two
calorimetric jets, each with energy greater than $10\GeV$, and to pass
stringent detector quality criteria for the vertexing, tracking and calorimetry.
A total of $1,248,350$ events are selected.  A cut on a $Z \rightarrow b\bar{b}$ 
event discriminant based on track DCA significances \cite{bbtag} yields a 
$b$-enriched sample of $176,980$ events.

To study the content of the selected data, a sample of 6 million hadronic Z 
decays have been generated with JETSET~7.4 \cite{Jetset}, and  passed through 
a GEANT based~\cite{Geant} simulation of the L3 detector. From this sample,
the $Z \rightarrow b\bar{b}$ event purity is determined to be 
$\pi_{b\bar{b}} = 0.828$.

\subsection{Selection of $\Bdstar \rightarrow \BorBstar\pi$ decays}

Secondary decay vertices and primary event vertices are reconstructed in three
dimensions by an iterative procedure such that a track can be a constituent of
no more than one of the vertices.  A calorimetric jet is selected
as a $\Bmeson$ candidate if it is one of the two most energetic jets in the
event, if a secondary decay vertex has been reconstructed from tracks associated
with that jet, and if the decay length of that vertex with respect to the event
primary vertex is greater than $3\sigma$, where $\sigma$ is the estimated error
of the measurement.

The decay of a $\Bdstar$ to a $\BorBstar$ meson and a pion is carried out
via a strong interaction and thus occurs at the primary event vertex.
In addition, the predicted masses for the $L=1$ states correspond to
relatively small $Q$ values, so that the decay pion ($\pi^{**}$) direction
is forward with respect to the $\Bmeson$ meson direction.  We take advantage of
these decay kinematics by requiring that, for each $\Bmeson$ meson candidate,
there is at least one track which is a constituent of the event primary vertex 
and which is located within 90 degrees of the jet axis.  A total of $60,205$ track-jet 
pairs satisfy these criteria.

To  decrease background, typically due to charged fragmentation
particles, only the track with the largest component of momentum in the direction
of the jet is selected.  This choice has been found~\cite{BoscCDF,BdstarALEPH}
to improve the purity of the signal.  The track is further required to have a transverse 
momentum with respect to the jet axis larger than $100\MeV$, to reduce 
background due to charged pions from $\Dstar \rightarrow \Dmeson\pi$ decays.
These selection criteria are satisfied by $48,022$ $\Bmeson\pi$ pairs with a $b$ 
hadron purity of $\pi_{\Bmeson} = 0.942$.

\subsubsection{B meson direction reconstruction}

The direction of the $\Bmeson$ candidate is estimated by taking a weighted
average in the $\theta$ (polar) and $\phi$ (azimuthal) coordinates
of directions defined by the vertices and by particles with a high rapidity
relative to the jet axis.  A numerical error-propagation method
\cite{Swain} makes it possible to obtain accurate estimates for the
uncertainty of the angular coordinates measured from vertex pairs.  These errors,
as well as the error for the decay length measurement used in the secondary
vertex selection, are calculated for each pair of vertices from the associated
error matrices.

Particles coming from the decay of $b$ hadrons produced in Z decays have
a characteristically high rapidity relative to the original direction of the
hadron when compared to that of particles coming from fragmentation.  A cut on
the particle rapidity distribution is thus a powerful
tool for selecting the $\Bmeson$ meson decay constituents\cite{BstarDELPHI}.
A second estimate for the direction of the $\Bmeson$ is obtained by summing the
momenta of all charged and neutral particles (excluding the $\pi^{**}$ candidate)
with rapidity $y > 1.6$ relative to the original jet axis.  Estimates for the
uncertainty of the coordinates obtained by this method are determined from
simulated $\Bmeson$ meson decays as an average value for all events.
The final $\Bmeson$ direction coordinates are taken as the error-weighted averages 
of these two sets of coordinates.

The resolution for each coordinate is parametrized by a two-Gaussian
fit to the difference between the reconstructed and generated values.  For
$\theta$, the two widths are $\sigma_1 = 18\mrad$ and $\sigma_2 = 34\mrad$ with
$68\%$ of the $\Bmeson$ mesons in the first Gaussian.  For $\phi$, the two widths
are $\sigma_1 = 12\mrad$ and $\sigma_2 = 34\mrad$ with $62\%$ of the $\Bmeson$
mesons in the first Gaussian.

\subsubsection{B meson energy reconstruction}

The energy of the $\Bmeson$ meson candidate is estimated by taking advantage of
the known center of mass energy at LEP to constrain the measured value.
The energy of the $\Bmeson$ meson from this method \cite{BdstarOPAL} can be
expressed as
\begin{equation}
  \label{eq:Benergy}
    E_{\Bmeson} = \frac{M^2_{\Zboson} + M^2_{\Bmeson} -
                        M^2_{\mathrm{recoil}}}{2M_{\Zboson}} \quad,
\end{equation}
where $M_{\Zboson}$ is the mass of the Z boson and $M_{\mathrm{recoil}}$ is the
mass of all particles in the event other than the $\Bmeson$.  To determine
$M_{\mathrm{recoil}}$, the energy and momenta of all particles in the event with
rapidity $y < 1.6$, including the $\pi^{**}$ candidate (regardless of its
rapidity), are summed and $M^2_{\mathrm{recoil}} = E^2_{y<1.6} - p^2_{y<1.6}$.
Fitting the difference between reconstructed and generated values for the B
meson energy with an asymmetric Gaussian yields a maximum width of $2.8\GeV$.


\section{Analysis of the $\boldBmeson\boldpi$ Mass Spectrum}

The combined $\Bmeson\pi$ mass is defined as
\begin{equation}
  \label{eq:BpiMass}
   M_{\Bmeson\pi} = \sqrt{M^2_{\Bmeson} + m^2_{\pi} + 2 E_{\Bmeson} E_{\pi} -
                          2 p_{\Bmeson} p_{\pi} cos\alpha} \quad,
\end{equation}
where $M_{\Bmeson}$ and $m_{\pi}$ are set to $5279\MeV$ and $139.6\MeV$,
respectively, and $\alpha$ is the measured angle between the $\Bmeson$ meson and
the $\pi^{**}$ candidate. The data mass spectrum is shown in Figure~\ref{fig:Voigt}.a 
together with the expected Monte Carlo background.

\subsection{Background function}

The background distribution is estimated from the Monte Carlo data sample,
excluding $\Bdstar\rightarrow\BorBstar\pi$ decays, and fitted with a six-parameter
threshold function given by
\begin{equation}
  \label{eq:BgdEqn}
  p_1 \times (x - p_2)^{p_3} \times e^{(p_4 \times (x - p_2) +
                                        p_5 \times (x - p_2)^2 +
                                        p_6 \times (x - p_2)^3)} \quad.
\end{equation}
Parameters $p_2$ through $p_6$ are fixed to the shape of the simulated background,
while the overall normalization factor $p_1$ is allowed to float freely in order to
obtain a correct estimate of the contribution of the background to the statistical
error of the signal.

\subsection{Signal function}

To examine the underlying structure of the signal, it is necessary
to unfold effects due to detector resolution.  The $\pi^{**}$ candidates are
expected to have typical momenta of a few $\GeV$.  In this range, the single
track momentum resolution is no more than a few percent with an angular
resolution better than $2\mrad$.  The dominant sources of uncertainty for the
mass measurement are thus the B meson angular and energy resolutions.  Monte
Carlo studies confirm that these two components are dominant and roughly
equal in magnitude.  This analysis thus concentrates on unfolding the effects
of these components by parametrizing and removing their contribution to the
mass resolution.

\subsubsection{Signal resolution and efficiency}

The dependence of the $\Bmeson\pi$ mass resolution and selection efficiency on
$Q$ value is studied by generating signal events at several different
values of $\Bdstar$ mass and Breit-Wigner width.  The simulated events are passed
through the same event reconstruction and selection as the data.  The resulting
$\Bmeson\pi$ mass distributions are each fitted with a Breit-Wigner function
convoluted with a Gaussian resolution (Voigt function) and the detector resolution
is extracted by fixing the Breit-Wigner width to the generated value.

The Gaussian width is found to increase linearly from $20\MeV$ to $60\MeV$
in the $\Bdstar$ mass range $5.6-5.8\GeV$.  This increase with $Q$ value is mainly
due to the angular component of the uncertainty, which increases
as a function of the opening angle $\alpha$.  The resolution is parametrized as
a linear function of the $\Bdstar$ mass from a fit to the extracted widths.
Similarly, the selection efficiency is found to increase slightly with $Q$ value
and the dependence is parametrized with a linear function.

Agreement between data and Monte Carlo for the $\Bmeson$ meson energy and
angular resolution is confirmed by analyzing $\Bstar\rightarrow\Bmeson\gamma$
decays selected from the same sample of $\Bmeson$ mesons.  The photon selection
for this test is the same as that described in reference \cite{BstarL3}.
A $\Bstar$ meson decays electromagnetically and hence has a negligible decay
width compared to the detector resolution.  As in the case of the $\Bmeson\pi$
mass resolution, the $\Bmeson$ meson energy and angular resolution are the
dominant components of the reconstructed $\Bmeson\gamma$ mass resolution.
Fits to the $M_{\Bmeson\gamma}-M_{\Bmeson}$ spectra are performed with the
combination of a Gaussian signal and the background function described above.
For the Monte Carlo, the Gaussian mean value is found to be
$M_{\Bmeson\gamma}-M_{\Bmeson} = (46.5 \pm 0.6\stat)\MeV$ with a width of
$\sigma = (11.1 \pm 0.7\stat)\MeV$.  The input generator mass
difference is $46.0\GeV$.  For the data, the Gaussian mean value is found to be
$M_{\Bmeson\gamma}-M_{\Bmeson} = (45.1 \pm 0.6\stat)\MeV$ with a width of
$\sigma = (10.7 \pm 0.6\stat)\MeV$.  Good agreement between the widths
of the data and Monte Carlo signals provides confidence that the $\Bmeson$
energy and angular resolution are well understood and simulated.

\subsubsection{Combined signal}

According to spin-parity rules,  five mass resonances are expected, corresponding to five
possible $\Bdstar$ decay modes: $\Btwostar\rightarrow\Bmeson\pi$,
$\Btwostar\rightarrow\Bstar\pi$, $\Bone\rightarrow\Bstar\pi$,
$\Bonestar\rightarrow\Bstar\pi$ and $\Bzerostar\rightarrow\Bmeson\pi$.  No
attempt is made to tag subsequent $\Bstar\rightarrow\Bmeson\gamma$ decays, as the
efficiency for selecting the soft photon is relatively low.  As a result, the
effective $\Bmeson\pi$ mass for a decay to a $\Bstar$ meson is shifted down by
the $46\MeV$ $\Bstar-\Bmeson$ mass difference.

\begin{figure}[htb]      
\begin{center}
\begin{tabular}{cc}
\epsfxsize 3.3 truein \epsfbox{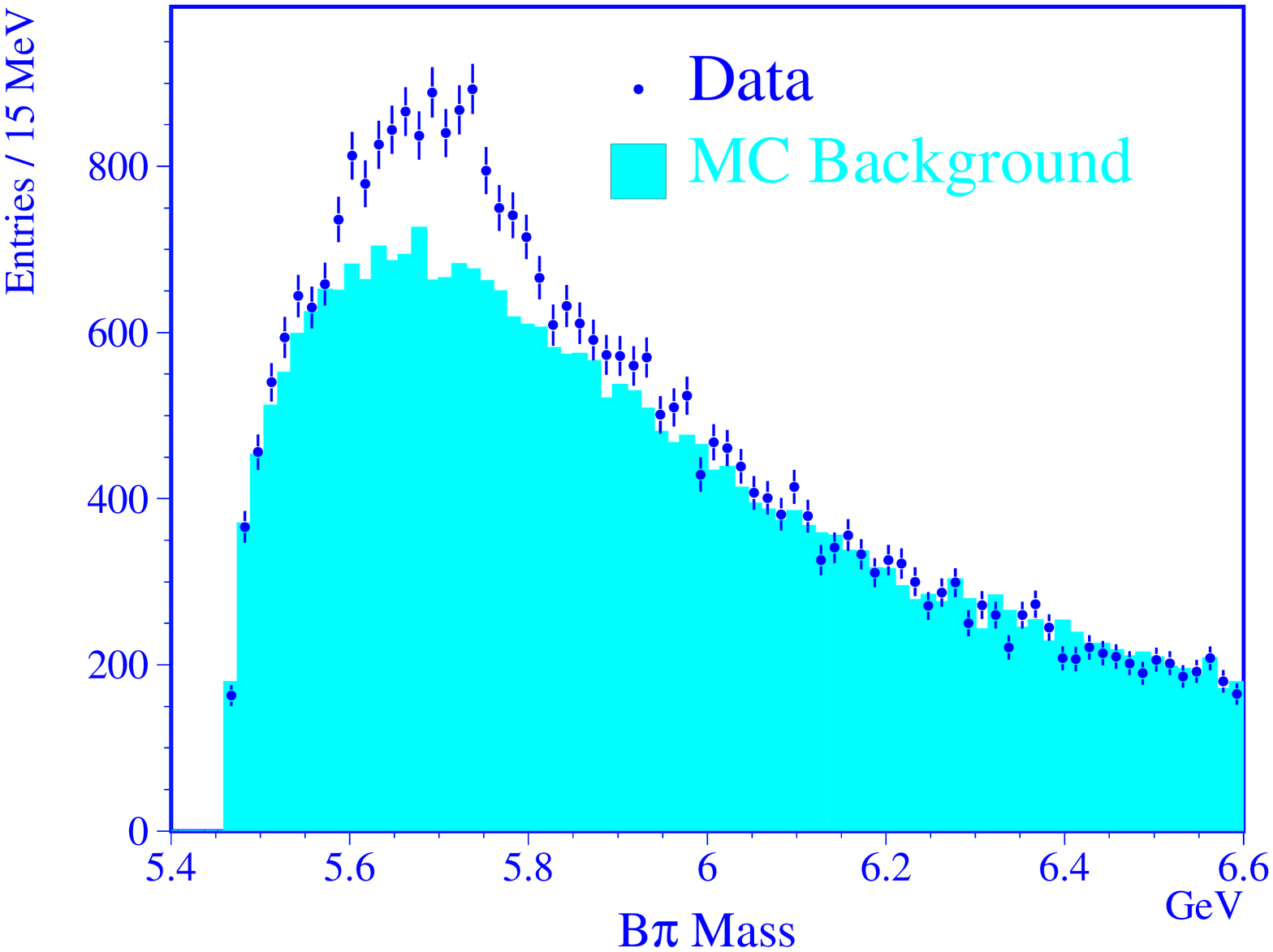} &
\epsfxsize 3.3 truein \epsfbox{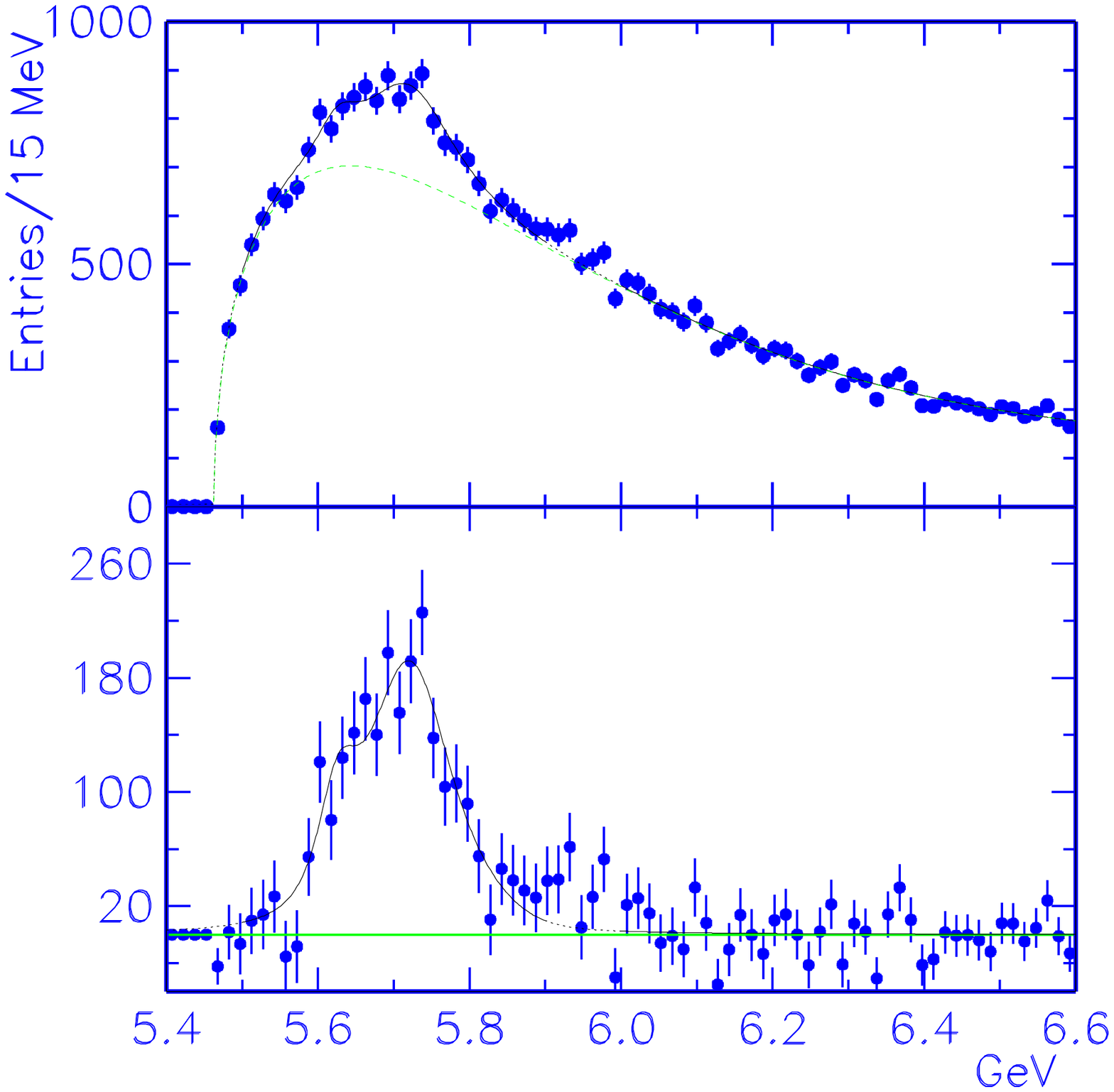}  \\
a) & b) 
\end{tabular}
\end{center}
\vskip -.2 cm
\caption[]{
\label{fig:Voigt}
\small a) Mass spectrum for selected  $\Bmeson\pi$ pairs. The dots are data and the 
shaded histogram represents the expected background from Monte Carlo normalized to
the sideband region $6.0-6.6 \GeV$;
b) fit to the data $\Bmeson\pi$ mass distribution with the five-peak
signal function and the background function described in the text.}
\end{figure}

The five resonances are fitted with five Voigt functions, with the relative
production fractions determined by spin counting rules.  The Gaussian
convolutions to the widths are determined by the resolution function.  Additional
physical constraints are applied to the mass differences and relative widths in
order to obtain the most information possible from the data sample.

Predictions for the mass differences $M_{\Btwostar} - M_{\Bone}$ and
$M_{\Bonestar} - M_{\Bzerostar}$ depend on several factors, including the $b$ and
$c$ quark masses and, in some cases, input from experimental data of the D meson
system.  The values are predicted to be roughly equal and in the range
$5-20\MeV$~\cite{Gronau,Eichten,Falk,Isgur,Ebert}.  We constrain both of the mass
differences to $12\MeV$. 

Predictions for the Breit-Wigner widths of the  $j_q = 3/2$ are extrapolated from 
measurements in the D meson system~\cite{Ddstar}  and are
expected to be roughly equal and about $20-25~\MeV$.  No precise predictions
exist for the $j_q=1/2$ states as there are no corresponding measurements in
the D system.  In general, however, they are also expected to be roughly equal,
although broader than those of the $j_q=3/2$ states.  We constrain
$\Gamma_{\Bone} = \Gamma_{\Btwostar}$ and
$\Gamma_{\Bzerostar} = \Gamma_{\Bonestar}$, but allow the widths of the
$\Btwostar$ and $\Bonestar$ to float freely in the fit.

\subsection{Fit results}

Monte Carlo events for each of the expected $\Bdstar$ decays are generated
and passed through the simulation and reconstruction programs and the
$\Bmeson\pi$ event selection.  The resulting mass spectra are combined with
background and fitted with the signal and background functions under the
constraints described above.  Mass values and decay widths for the $\Btwostar$
and $\Bonestar$ resonances and the overall normalization are extracted from the
fit and found to agree well with the generated values.  All differences lie
within the statistical error and have no systematic trend.

The data $\Bmeson\pi$ mass spectrum is fitted with the combined signal and
background functions, allowing the normalization parameters to float freely.
The resulting fit, shown in Figure~\ref{fig:Voigt}, has a $\chi^2$ of $39$ for
$74$ degrees of freedom.  A total of $2652$ events occupy the signal region
corresponding to a relative $\BdstarUD$ production rate of
$\sigma(\BdstarUD)/\sigma(\BmesonUD) = 0.39 \pm 0.05\stat$.  The mass and width
of the $\Btwostar$ are found to be $M_{\Btwostar} = (5770 \pm 6\stat)\MeV$ and
$\Gamma_{\Btwostar} = (21 \pm 24\stat)\MeV$ and the mass and width of the
$\Bonestar$ are found to be $M_{\Bonestar} = (5675 \pm 12\stat)\MeV$ and
$\Gamma_{\Bonestar} = (75 \pm 28\stat)\MeV$.

\subsection{Systematic uncertainty}

Sources of systematic uncertainty and their estimated contributions to the
errors of the measured values are summarized in Table~\ref{tab:syst}.  The
$b$ hadron purity of the sample is varied from $91\%$ to $96\%$.  The fraction
of $b$ quarks hadronizing to $\BmesonUD$ mesons is taken to be $79\%$ and is
varied between $74\%$ and $83\%$ in accordance with the recommendations of the
LEP $\Bmeson$ Oscillation Working Group \cite{LEPBoscWG}.  These variations
effect only the overall $\Bdstar$ production fraction.

Systematic effects due to background modelling are studied by varying the shape
parameters of the background function and by performing the fit with other
background functions to study the effect on the measured values.  Contributions
to the error due to modelling of the signal are estimated for the mass and width
constraints: the $M_{\Btwostar}-M_{\Bone}$ and $M_{\Bonestar}-M_{\Bzerostar}$
mass differences are varied in the range $6-18\MeV$ and the
$\Gamma_{\Bone}/\Gamma_{\Btwostar}$ and $\Gamma_{\Bzerostar}/\Gamma_{\Bonestar}$
ratios are varied between $0.8$ and $1$.  Effects due to uncertainty in the
resolution and efficiency functions are estimated by varying the slopes and
offsets of the linear parametrizations.

Three-body decays of the type $\Btwostar\rightarrow\Bmeson\pi\pi$ have been
generated and passed through the simulation and reconstruction programs and the
$\Bmeson\pi$ event selection.  $\Bmeson\pi$ pairs, for which only one of the pions
is tagged, are studied as a possible source of resonant background.  The resulting
reflection is found to contribute insignificantly to the background in regions of
small $Q$ value.  Similarly, generated $\BdstarS\rightarrow\Bmeson\Kmeson$ decays,
where the $\Kmeson$ is mistaken for a $\pi$ are found to contribute only slightly
to the low $Q$ value region and their effects are included in the background
modelling uncertainty contribution.


\section{Conclusion}

We measure for the first time the masses and decay widths of the $\Btwostar$
($j_q=3/2$) and $\Bonestar$ ($j_q=1/2$) mesons.  From a constrained fit to the
$\Bmeson\pi$ mass spectrum, we find
\begin{eqnarray*}
  M_{\Btwostar}      & = & (5770 \pm 6 \stat \pm 4 \syst)\MeV \\
  \Gamma_{\Btwostar} & = & (23 \pm 26 \stat \pm 15 \syst)\MeV \\
  M_{\Bonestar}      & = & (5675 \pm 12 \stat \pm 4 \syst)\MeV \\
  \Gamma_{\Bonestar} & = & (76 \pm 28 \stat \pm 15 \syst)\MeV \quad .
\end{eqnarray*}
The relative $\BdstarUD$ production rate, including all $L=1$ spin states,
is measured to be
\begin{eqnarray*}
  \frac{\Br(b\rightarrow\BdstarUD\rightarrow\BorBstar\pi)}
       {\Br(b\rightarrow\BmesonUD)} = 0.39 \pm 0.05 \stat \pm 0.06 \syst
\end{eqnarray*}
where isospin symmetry is employed to account for decays to neutral pions.

\begin{table}[tb]
  \begin{center}
    \caption{Sources of systematic uncertainty and their estimated contributions
             to the errors of the measured values.\label{tab:syst}}
    \vspace{0.2cm}
    \begin{tabular}{l|r|r|r|r|r}
      {\bf Sources} &
      {\bf $M_{\Btwostar}$} & {\bf $\Gamma_{\Btwostar}$} &
      {\bf $M_{\Bonestar}$} & {\bf $\Gamma_{\Bonestar}$} &
      {\bf $f^{**}$} \\
      \hline
      $b$ purity
      & ---     & ---      & ---      & ---      & $\pm 0.02$ \\
      $\BmesonUD$ fraction
      & ---     & ---      & ---      & ---      & $\pm 0.03$ \\
      background
      & $\pm 2$ & $\pm  9$ & $\pm  3$ & $\pm 9$  & $\pm 0.05$ \\
      M constraints
      & $\pm 3$ & $\pm  7$ & $\pm  3$ & $\pm 7$  & $<0.01$    \\
      $\Gamma$ constraints
      & $<1$    & $\pm  2$ & $<1$     & $\pm 2$  & $<0.01$    \\
      resolution
      & $\pm 2$ & $\pm  9$ & $\pm  1$ & $\pm 9$  & $\pm 0.01$ \\
      efficiency
      & $<1$    & $\pm  1$ & $<1$     & $\pm 1$  & $\pm 0.01$ \\
      \hline
      {\bf Total}
      & $\pm 4$ & $\pm 15$ & $\pm  4$ & $\pm 15$ & $\pm 0.06$ \\
    \end{tabular}
  \end{center}
\end{table}

\section*{Acknowledgments}

I wish to thank my colleagues Steven Goldfarb and Franz Muheim for their help in 
preparing this talk.

\end{document}